\documentstyle[12pt]{article}
\begin{document}

\begin{flushright}
FERMILAB-Pub-95-374-A \\
DART-HEP-95/07\\
December 1995
\end{flushright}

\vspace{1in}
\begin{center}
{\Large{\bf  Modeling Thermal Fluctuations: Phase Mixing and Percolation}}\\
\vspace{.5in}

{\bf Marcelo Gleiser }\\
{\em Department of Physics and Astronomy, Dartmouth College,}\\
{\em Hanover, NH 03755, USA }\\{\em email: gleiser@peterpan.dartmouth.edu}
\vspace{.2in}

{\bf Andrew F. Heckler} and {\bf Edward W. Kolb}\\
{\em  NASA/Fermilab Astrophysics Center,
Fermi National Accelerator Laboratory,}\\
{\em Batavia, IL 60510, USA}\\{\em email: aheckler@fnas04.fnal.gov,
rocky@rigoletto.fnal.gov}

\end{center}

\begin{abstract}
We consider the nonequilibrium dynamics of a
a real scalar field in a degenerate double-well potential. The system is
prepared in the lowest free energy state in one of the wells and the dynamics
is driven by the coupling of the field to a thermal bath. Using a simple
analytical model, based on the subcritical bubbles method, we compute
the fraction of the total volume which fluctuates to the opposite phase as a
function of the parameters of the potential. Furthermore, we show how
complete phase mixing, {\em i.e.} symmetry restoration, is related to
percolation, which is
dynamically driven by domain instability. Our method describes
quantitatively recent results obtained by numerical simulations, and is
applicable to systems in the Ising universality class.
\end{abstract}

\vspace{.75in}
submitted to {\em Physical Review Letters}.

\newpage

It is well-known that an Ising ferromagnet loses its net magnetization
above a certain temperature. The simplest mean-field
description of this phenomenon is
based on the Ginzburg-Landau model, in which the average
magnetization $M=(V^{-1})
\int dV~M(\bf x)$
is the order
parameter, and the thermodynamic potential is expanded to fourth power in
$M$, $V(M) = a(T-T_c)M^2 + bM^4$ \cite{Landau}. We are only interested in
the case of vanishing external magnetic field.
Above $T_c$, the lowest free energy state has zero net magnetization,
while below $T_c$ the material acquires a magnetization $M= \pm [a|T-T_c|/
2b]^{1/2}$.

It is also well known that mean-field theory breaks down close to $T_c$;
for example, although mean-field theory correctly
predicts that the correlation length
diverges as $\xi\sim |(T-T_c)/T_c|^{-\nu}$, it gives the value of the critical
exponent $\nu=0.5$, while numerical simulations find $\nu=0.630(2)$
\cite{Goldenfeld}. In order to handle the infrared divergences that appear
near $T_c$, the renormalization group is used to relate a given theory to
an equivalent theory with smaller correlation lengths and thus better
behaved in the infrared. With the $\varepsilon$-expansion, one works in
$4-\varepsilon$ dimensions and finds a fixed point of order $\varepsilon$
of the renormalization group equations, taking the limit $\varepsilon
\rightarrow 1$ in the end. To second order in $\varepsilon$,
$\nu = {1\over 2} +{1\over {12}}\varepsilon + {7\over {162}}\varepsilon^2
\simeq 0.63$, a remarkable result \cite{Brezin}.

In the present work, we would like to follow a somewhat orthogonal approach
to study a closely related question. Consider again an Ising ferromagnet in the
absence of an external magnetic field,
which is well below its critical temperature and has been prepared with
all spins initially pointing in one
direction. In the thermodynamic limit, this is the broken-symmetric state.
As the temperature is increased, thermal fluctuations will flip groups of
spins, so that the absolute value of the magnetization will start decreasing.
At some temperature $T_c^*<T_c$, $|M|\rightarrow 0$. (This is another way
of expressing the breakdown of mean-field theory.)
The question we would like to address is the following: At a given temperature
below $T_c$, what is the fraction of the volume occupied by each of the two
possible phases of the system as a function of the parameters of the potential?
Can we provide a method for computing
this fraction which somehow encompasses the
breakdown of mean-field theory, without resorting to the renormalization
group?

We start by writing the homogeneous part of the free-energy density as
\begin{equation}
V(\phi ,T)={a\over 2}\left (T^2-T_2^2\right )\phi^2-
{{\alpha}\over 3}T\phi^3 +{{\lambda}\over 4}
\phi^4 \:.
\label{freen}
\end{equation}
It is easy to see that, with a simple field redefinition,
we can rewrite $V(\phi ,T)$ as a Ginzburg-Landau potential with a
temperature-dependent magnetic field. At the critical temperature, this field
vanishes and we recover the degenerate double-well potential.

Introducing dimensionless variables ${\tilde x} = a^{1/2} T_2
x$, ${\tilde t} = a^{1/2} T_2 t$, $X = a^{-{1/4}}
T_2^{-1} \phi$, and $\theta = T/T_2$, the Hamiltonian is,
\begin{eqnarray}
{{H[X]}\over {\theta}} &=& {1\over {\theta}}\int d^3{\tilde x}\left [
{1\over 2}\left| {\tilde \bigtriangledown} X\right|^2 +
{1\over 2}\left (\theta^2 -1
\right )X^2 \right. \nonumber \\
& & \left. -{{{\tilde \alpha}}\over 3}\theta X^3+
{{{\tilde \lambda}}\over 4}X^4\right ] \:,
\label{hamilton}
\end{eqnarray}
where ${\tilde \alpha} = a^{-{3/4}} \alpha$, and ${\tilde
\lambda} = a^{-{1/2}} \lambda$ (henceforth we drop the
tildes). At the critical temperature
$\theta_c=(1-2\alpha^2/9\lambda)^{-1/2}$ the two minima,
at $X_0=0$ and $X_+=
{{\alpha\theta}\over {2\lambda}}\left [ 1+
\sqrt{1-4\lambda\left (1-1/\theta^2\right )/\alpha^2}\right ]$,
are degenerate. In what
follows, we are only interested in the system at $\theta_c$.

In a recent work, Borrill and Gleiser (BG) simulated the dynamics of the above
system coupled to a Markovian thermal bath \cite{BG}. (This is why we
wrote the potential as in Eq.~\ref{freen}.)
In analogy with Ising ferromagnetism,
it is useful to think of the system described by the Hamiltonian above as
having two phases, the ``0 phase'', for $X<X_{\rm max}$, and the ``+ phase'',
for
$X\geq X_{\rm max}$, where $X_{\rm max}$ is at the maximum of the double-well
potential.
The initial
conditions were chosen so that the system started in the 0 phase.
Thus, $V_0^t/V_T(t=0)=1$, where $V_0^t$ is the total volume in the
0 phase, $V_T$ is the constant total volume, and $V_0^t/V_T(t)+V_+^t/V_T(t)=1$.
The coupling of this system to a thermal bath will induce the nucleation of
domains of + phase. For a small free-energy barrier between the two phases,
domains of the 0 phase will also nucleate
within domains of + phase, and so on,
resulting in a very complicated domain structure.
It is necessary to distinguish between
two possible kind of domains, connected
and disconnected. {\it Connected domains}
percolate throughout the volume, and thus
cannot be surrounded by a domain of the opposite phase. {\it Disconnected
domains} have finite volume and are always surrounded by the other phase.
In general, $V_{0(+)}^t = V_{0(+)}^c + V_{0(+)}^d$.

Keeping the system
at $\theta_c$ and fixing $\alpha=0.065$ (this seemingly {\it ad hoc}
choice was inspired by the electroweak effective potential, although any other
value would do), BG measured the equilibrium fractions, $f_0\equiv V_0^t/V_T$
and $f_+\equiv V_+^t/V_T$, as a function of the coupling $\lambda$. They
found that for $\lambda \geq \lambda_c\simeq 0.025$, $f_0=f_+=0.5$, that is,
the two phases completely mix (see Fig. ~1). In other words, for $\lambda\geq
\lambda_c$,
the symmetry is restored, even though the mean-field potential still has
a double-well shape. Clearly, for $\lambda \geq \lambda_c$, mean-field theory
breaks down, and the system is better described by an effective free-energy
density
with a single minimum at $X_{\rm max}=0$. This situation is exactly analogous
to an Ising ferromagnet for $T\geq T_c^*$.

In what follows, we will reproduce these results with a simple statistical
model for the thermal fluctuations. Our approach is completely general,
in that it can be easily adapted to other systems described by a similar
double-well potential, {\it i.e.}, for systems in the universality class
of the Ising ferromagnet.
We will assume that the large-amplitude
fluctuations from the 0(+) phase into the +(0) phase can be modelled by
spherically-symmetric subcritical
bubbles of Gaussian shape of a given radius and amplitude (For previous
treatments of subcritical bubbles see Ref. \cite{SUBBUB}.),
\begin{equation}
\phi_c(r)=\phi_ce^{-r^2/R^2}~,~~~\phi_0(r)=\phi_c\left (1-
e^{-r^2/R^2}\right ) \:,
\end{equation}
where $R$ is the radial size of the configuration, and $\phi_c$ is the
value of the field amplitude at the bubble's core,
away from (and into)
the 0 phase. For these configurations to interpolate between the two phases in
the system, $\phi_c\geq \phi_{\rm max}$. With this {\it ansatz}, the free
energy of the fluctuations assumes the general form,
$F_{\rm sc}(R,\phi_c,T)= b\phi_c^2R + c\beta(\phi_c,T)R^3$,
where $b$, and $c$ are numerical constants and $\beta(\phi_c,T)$ is a
polynomial which depends on the particular potential used. We will further
assume that the nucleation rate for these configurations is obtained from a
Gibbs distribution, $G(\phi,R)=A{\rm exp}\left [-F_{\rm sc}/T\right ]$, where
$A$ is a constant. The nucleation rate per unit volume is then
$\Gamma = \int G d\phi~dR$.

The number density of bubbles of, say, the + phase
with radii between $R$ and $R+\delta R$
and amplitudes between $\phi$ and $\phi+\delta \phi$ at time $t+\delta t$ is
\begin{eqnarray*}
n_+(R &+& \delta R,\phi  + \delta \phi,t+\delta t)-n_+(R,\phi+\delta \phi,
t+\delta t) \\
& -&\left [ n_+(R+\delta R,\phi,t+\delta t)
- n_+(R,\phi,t+\delta t)\right ]~. \\
\end{eqnarray*}
Changes in the number density are generated by three different processes:
i) bubbles can shrink into and out of this interval. [We assume the time
dependence of the amplitude is closely related to the radial time-dependence,
as recent numerical studies have demonstrated \cite{OSCILLONS}.]; ii)
bubbles can be thermally nucleated into this interval; iii) bubbles
can be thermally destroyed out of this interval by inverse nucleation.
Expanding to first order in $\delta R,~\delta \phi$,
and $\delta t$, we obtain a
Boltzmann equation for the bubble distribution function $f_+^d(R,\phi,t)$,
\begin{eqnarray}\label{boltz}
\frac{\partial f^d_+(R,\phi,t)}{\partial t} &=&
-|v|\frac{\partial f^d_+}{\partial R} + {{V_0^c}\over {V_T}}
G_{0\rightarrow +} \nonumber \\
& & - {{V_+^d}\over {V_T}} G_{+\rightarrow 0}~,
\end{eqnarray}
where the bubble density distribution function for domains of
the + phase, [hence the
superscript $d$, for disconnected], is defined as
\begin{equation}
f^d_+(R,\phi,t)\equiv {{\partial^2 n_+}\over {\partial\phi \partial R}}~,
\end{equation}
and
$G_{0(+)\rightarrow +(0)}$ is the Gibbs distribution for the nucleation rate
per unit volume
of bubbles of the +(0) phase within the 0(+) phase. Note that we have written
the Boltzmann equation to be consistent with the initial conditions used in
the simulations, so that only disconnected
domains contribute to the fraction of
the volume in the + phase. It is straightforward
to adapt the equation to different
initial conditions.

In order to proceed, we note that the total fraction of volume in the + phase
can be written as
\begin{equation}\label{gamma}
\gamma \simeq \int_{\phi_{\rm max}}^{\infty}\int_{R_{\rm min}}^{\infty}
\left ({{4\pi R^3}\over 3}\right ){{\partial^2 n_+}\over {\partial \phi
\partial R}}
d\phi dR~~.
\end{equation}
For a degenerate double-well, $G_{+\rightarrow 0}=G_{0\rightarrow +}
\equiv G$, and the Boltzmann equation can be written as,
\begin{eqnarray}\label{boltz2}
\frac{\partial f^d_+(R,\phi,t)}{\partial t} &=&
-|v|\frac{\partial f^d_+}{\partial R} + (1-2\gamma)G~.
\end{eqnarray}

Imposing the physical condition $f^d_+(R\rightarrow \infty,\phi,t)
\rightarrow 0$, we can solve for the equilibrium distribution function
($\dot f^d_+=0$), and use it to compute the equilibrium
fraction of the volume in the + phase, $\gamma_{\rm eq}$, which is the
quantity measured in the BG simulation. The general solution is,
\begin{equation}\label{gamma2}
\gamma_{\rm eq} = {{I(\phi_{\rm max},R_{\rm min})}\over
{1+2I(\phi_{\rm max},R_{\rm min})}}~,
\end{equation}
where,
\begin{eqnarray}
I(\phi_{\rm max},R_{\rm min})&=&{A\over {|v|}}\int_{R}^{\infty}
\int_{R_{\rm min}}^{\infty}
\int_{\phi_{\rm max}}^{\infty} dR'dRd\phi~ \nonumber \\
& & \times {{4\pi}\over 3}R^3{\rm exp}\left [
-F_{\rm sc}(R',\phi)\right ]~,
\end{eqnarray}
and $\phi_{\rm max}$ is at the maximum of the double well potential, and the
minimum radius  is taken to be the lattice spacing in the BG simulation, {\em
i.e.} $R_{\rm min} = 1$, which sets the coarse-graining scale. An
analytical expression for $I(\phi_{\rm max},R_{\rm min})$
can be obtained if we
write $F_{\rm sc}(R',\phi)\approx b\phi^2R'$. This is a good approximation in
the case that fluctuations are small enough that the volume term is
sub-dominant.

In Fig. 1, we show $\gamma_{\rm eq}$ as a function of $\lambda$. The dots are
the results of BG, while the curves are the results of the integration of the
Boltzmann equation, for different values of the single parameter $A/|v|$. It is
clear that for $\lambda{\
\lower-1.2pt\vbox{\hbox{\rlap{$<$}\lower5pt\vbox{\hbox{$\sim$}}}}\ }
\lambda_c$, we obtain an excellent fit to the
data with $A/|v|=60$.

For larger values of $\lambda$, our method
underestimates the fraction of volume in the + phase. This can be understood
by noting that our kinetic description does not include possible terms
which account for the coalescence of nearby domains. As fluctuations become
more probable, these terms will play an increasingly important r\^ole. In fact,
it is easy to understand how the onset of domain instability is intimately
related to percolation of the + phase. (An excellent introduction to
percolation
theory can be found in Ref. \cite{Percolation}.)

Consider a large spherical domain of the + phase
of radius $R$. There are three ways by
which the volume $V$ of this domain can change; i) it may shrink by surface
tension with a velocity $v$; ii) a small bubble of the + phase may nucleate
just outside it; iii) a small bubble of the 0 phase may nucleate inside it.
Assuming that the bubble of
the 0 phase nucleates just inside the large domain (see Fig. 2), we can write
an approximate equation for the change in the volume $V$,
\begin{equation}
{{dV}\over {dt}}= -v4\pi R^2 +\left (\Gamma_{0\rightarrow +}\Delta V\right )
{{4\pi}\over 3}r^3 - \left (\Gamma_{+\rightarrow 0}\Delta V'\right )
{{4\pi}\over 3}r'^{3}~,
\end{equation}
where $\Gamma_{0(+)\rightarrow +(0)}\Delta V^{(')}$
is the nucleation rate for a bubble of the +(0)-phase of radius $r^{(')}$ in
the neighborhood of
the domain wall. Assuming for simplicity that $r=r'$, and recognizing that
$\Gamma_{+\rightarrow 0}(r) = \Gamma_{0\rightarrow +}(r) $ for a degenerate
double-well potential,
the condition for domain instability, ${{dV}\over {dt}}> 0$, becomes,
\begin{equation}\label{instability}
\Gamma r^4 > {3\over {8\pi}}\left ({R\over r}\right )v~.
\end{equation}

On the other hand, in order for the + phase to percolate,
$\gamma_{\rm eq}>p_c$, where $p_c$ is the critical percolation probability.
Using Boltzmann's equation, we can approximately write
\begin{equation}\label{percolation}
\gamma_{\rm eq}= {g\over {1+2g}},~~g\simeq {{4\pi}\over {3v}}\Gamma r^4~.
\end{equation}
Thus, for percolation, we obtain the inequality,
\begin{equation}\label{percolation2}
\Gamma r^4>{3\over {4\pi}}{{p_c}\over {1-2p_c}}v~.
\end{equation}
For a simple cubic lattice, $p_c=0.311$. Note that this is remarkably close to
the value of $\gamma_{\rm eq}$ for $\lambda_c$, the point where the kinetic
description breaks down (see Fig. 1).

Moreover, comparing the two inequalities of Eqs.~
\ref{instability} and \ref{percolation2}, it is clear that they are
satisfied at similar values of $\Gamma r^4$. In particular, writing for
simplicity $p_c=1/3$, they are equal for $R=2r$. This simple argument
strongly suggests that the
onset of percolation is dynamically driven by
the nucleation of small bubbles in the neighborhood of large domains.

As stated before, once the barrier between the phases is small enough, the
domain wall becomes unstable and nucleation of small bubbles is the dominant
process for changing a volume of space from one
phase to another. The
equations determining the fraction of volume in each phase  can then be
approximated as
\begin{eqnarray}\label{}
\frac{d V_{+}}{dt} &=& \left(
{{4\pi}\over 3}r^3\Gamma_{0\rightarrow +}\right) V_{0}(t) -
\left( {{4\pi}\over
3}r^{3}\Gamma_{+\rightarrow 0}\right) V_{+}(t) \nonumber \\
V_{0}(t) &=& V_{\rm tot} - V_{+}(t)\, ,
\end{eqnarray}
where $r$ is the average radius of a nucleated bubble.  The solution to these
equations is
\begin{eqnarray}\label{percsolution}
\frac{V_{+}(t)}{V_{\rm tot}} = \frac{1}{2} +
\left(\frac{V_{+}(0)}{V_{\rm tot}}
- \frac{1}{2}  \right) e^{-(8\pi\Gamma r^3/3)t}.
\end{eqnarray}
Notice that the equilibrium value $V_{+}(\infty)/V_{\rm tot} = \gamma_{\rm eq}
= 1/2$, independently of $V_{+}(0)$.

Therefore, once the  domain wall becomes unstable, the system will equilibrate
to $\gamma = 1/2$ in a time scale $\sim (\Gamma r^3)^{-1}$,  which, at
percolation,
is of order the light
crossing time $r$ (see Eq.~\ref{percolation2}).
That is, once the
system percolates, the stable, equilibrium, mean value of the field is at
$\phi_{\rm max}$, which is the value exactly between  the two minima.

As shown eq. (\ref{percsolution}), even if the system is perturbed away from
$\gamma = 1/2$, it will quickly relax back to this value. This indicates that
the percolation point  {\em is}  the point of symmetry restoration: the
symmetry of the true (coarse-grained) effective potential has been
restored, even
though the mean field potential still describes a double-well potential. It is
important to note, however, that even if the average value of the field is
$\phi_{\rm max}$, there are widespread large amplitude
fluctuations with average volume set by the correlation length;
the system is far from being locally homogeneous.

We conclude that our method based on subcritical bubbles gives a quantitatively
accurate description of the behavior of thermal fluctuations for
models in the Ising universality class. It also provides a dynamical
picture of symmetry restoration, and the breakdown of mean-field theory.
Furthermore, the method relates the breakdown of mean-field theory to a
critical
value of a given parameter, which is easily calculable. For values of the
parameter larger than the critical value,
large domains become unstable
to growth due to the nucleation of nearby bubbles, percolation ensues, and
symmetry or, more accurately, complete phase mixing, is restored.

MG was partially supported at Dartmouth by the National Science Foundation
through a  Presidential Faculty Fellows
Award no. PHY-9453431 and by a NASA grant no. NAGW-4270. AFH and EWK were
supported at Fermilab by both the DOE and NASA (NAG5-2788).


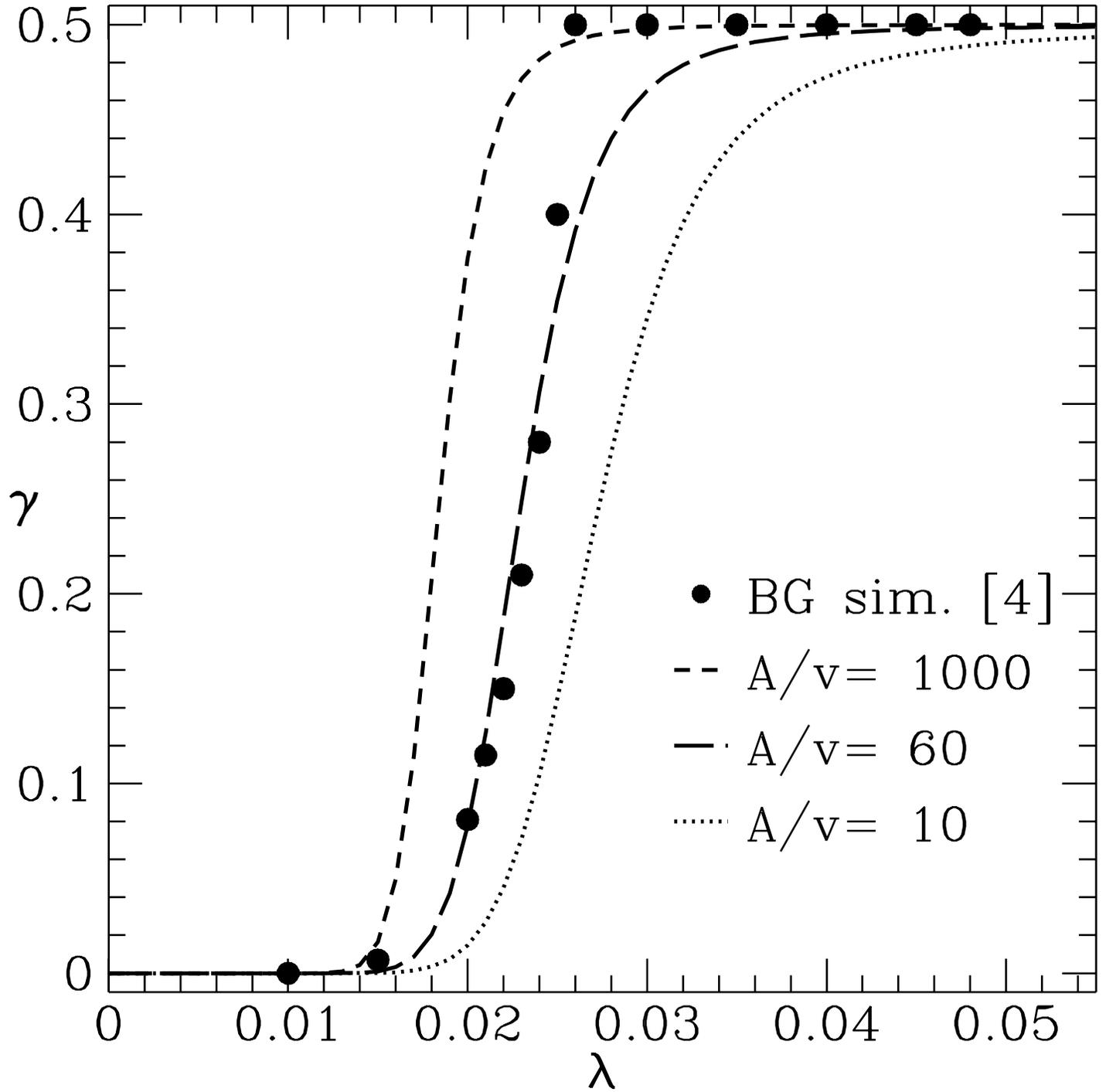
Figure 1. The fraction of the volume in the + phase. The dots are from the
numerical simulations of BG, while the lines are the solutions of the Boltzmann
equation for different values of the parameter $A/|v|$.

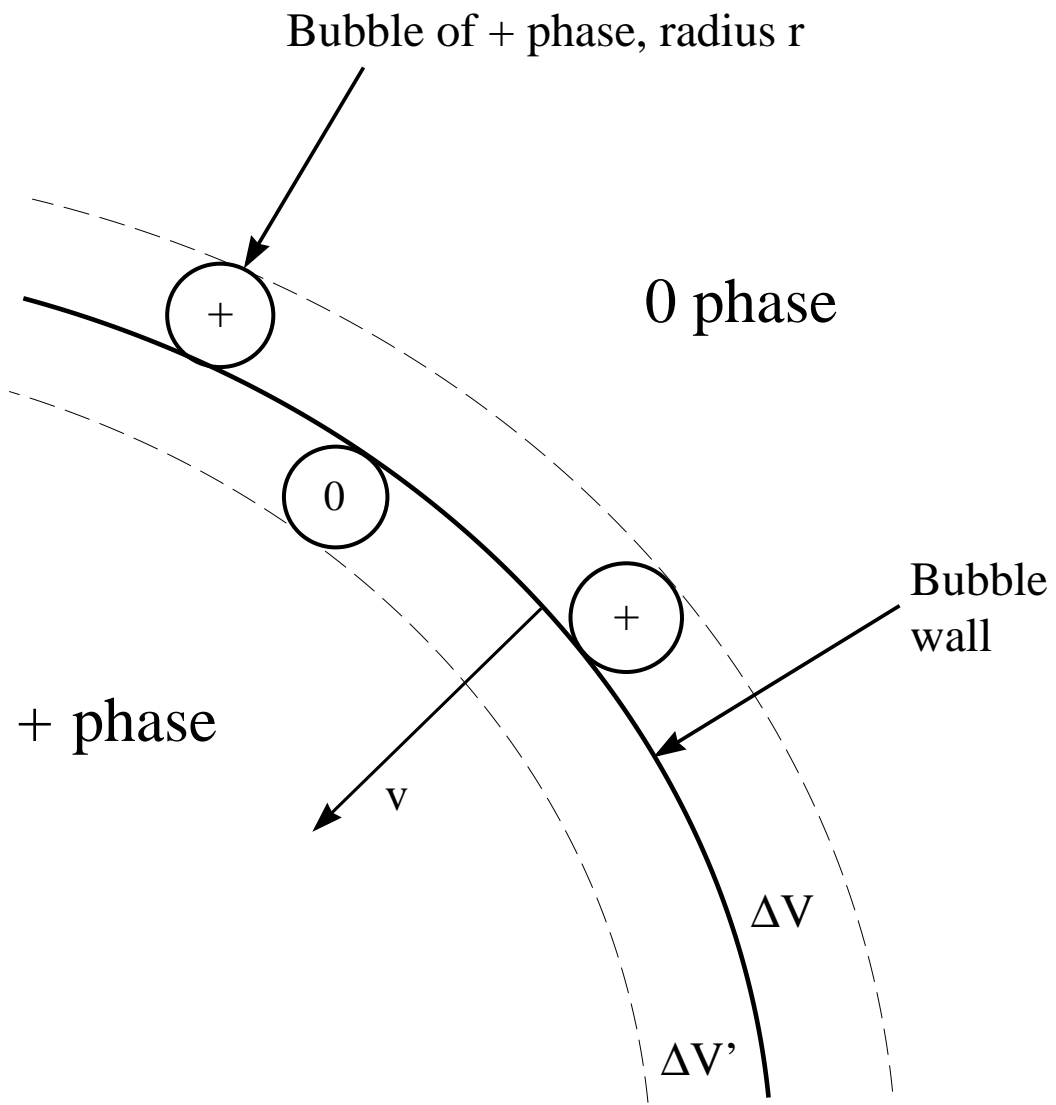
Figure 2. Schematic of domain instability. The surface tension on bubble wall
will tend to shrink the large bubble with wall velocity $v$. However, more
bubbles of + phase will nucleate just outside the large bubble wall than
bubbles of 0 phase just inside the wall because $\Delta V > \Delta V'$
(nucleation rates of 0 and + bubbles are equal). This will tend to make the
large bubble grow. When the barrier between the phases is small enough,
nucleation dominates over bubble shrinking, causing the wall to become
unstable to rapid growth.

\end{document}